\documentclass[fleqn,usenatbib]{mnras}

\usepackage{newtxtext,newtxmath}
\usepackage[T1]{fontenc}
\usepackage{ae,aecompl}

\usepackage{subfig,graphicx}	% Including figure files
\usepackage{amsmath}	% Advanced maths commands
\usepackage{wasysym}
\usepackage{enumitem}
\usepackage{footmisc}
\usepackage{xcolor}
\usepackage[capitalise]{cleveref}
\usepackage{hyperref}
\hypersetup{draft}
\usepackage{todonotes}
\usepackage{multicol}% http://ctan.org/pkg/multicols

\title[X-ray AGN in massive galaxy clusters]{
The environmental dependence of X-ray AGN activity at $z\sim0.4$}

 \author[E. Noordeh et al.]{
E. Noordeh,$^{1,2\thanks{E-mail: emiln@stanford.edu}}$
R.E.A. Canning,$^{1,2}$
A. King,$^{1,2}$
S.W. Allen,$^{1,2,3}$
A. Mantz,$^{1,2}$
\newauthor~R.G. Morris,$^{1,2,3}$
S. Ehlert,$^{4}$
A. von der Linden,$^{5}$
W.N. Brandt,$^{6,7,8}$
B. Luo,$^{9}$
\newauthor~Y. Q. Xue,$^{10}$
P. Kelly$^{11}$
 \\
 $^{1}$Department of Physics, Stanford University, 382 Via Pueblo Mall, Stanford, CA 94305-4060, USA\\
 $^{2}$Kavli Institute for Particle Astrophysics and Cosmology, Stanford University, 452 Lomita Mall, Stanford, CA 94305-4085, USA\\
 $^{3}$SLAC National Accelerator Laboratory, 2575 Sand Hill Road, Menlo Park, CA 94025, USA\\
 $^{4}$Marshall Space Flight Center, Huntsville, AL 35812 USA\\
 $^{5}$Department of Physics and Astronomy, Stony Brook University, Stony Brook, NY 11794, USA\\
 $^{6}$Department of Astronomy \& Astrophysics, The Pennsylvania State University, University Park, PA 16802, USA\\
 $^{7}$Institute for Gravitation and the Cosmos, Department of Physics, Pennsylvania State University, University Park, PA 16802, USA\\
 $^{8}$Department of Physics, 104 Davey Laboratory, The Pennsylvania State University, University Park, PA 16802, USA\\
 $^{9}$School of Astronomy and Space Science, Nanjing University, Nanjing 210093, China\\
 $^{10}$CAS Key Laboratory for Research in Galaxies and Cosmology, Department of Astronomy, \\
 University of Science and Technology of China, Hefei 230026, China\\
 $^{11}$Minnesota Institute for Astrophysics, University of Minnesota, 115 Union St. SE, Minneapolis, MN 55455, USA
 }

\date{Accepted XXX. Received YYY; in original form ZZZ}

\pubyear{2020}

\begin{document}
\label{firstpage}
\pagerange{\pageref{firstpage}--\pageref{lastpage}}
\maketitle

% Abstract of the paper
\begin{abstract}
We present an analysis of the X-ray Active Galactic Nucleus (AGN) population in a sample of seven massive galaxy clusters in the redshift range $0.35<z<0.45$. We utilize high-quality {\it Chandra} X-ray imaging to robustly identify AGN and precisely determine cluster masses and centroids. Follow-up VIMOS optical spectroscopy allows us to determine which AGN are cluster members. Studying the subset of AGN with 0.5-8 keV luminosities $>6.8\times10^{42}~\mathrm{erg~s^{-1}}$, within $r\leq2r_{500}$ (approximately the virial radius), we find that the cluster AGN space density scales with cluster mass as $\sim M^{-2.0^{+0.8}_{-0.9}}$. This result rules out zero mass dependence of the cluster X-ray AGN space density at the 2.5$\sigma$ level. We compare our cluster X-ray AGN sample to a control field with identical selection and find that the cluster AGN fraction is significantly suppressed relative to the field when considering the brightest galaxies with $V<21.5$. For fainter galaxies, this difference is not present. Comparing the X-ray hardness ratios of cluster member AGN to those in the control field, we find no evidence for enhanced X-ray obscuration of cluster member AGN. Lastly, we see tentative evidence that disturbed cluster environments may contribute to enhanced AGN activity.
\end{abstract}

% Select between one and six entries from the list of approved keywords.
% Don't make up new ones.
\begin{keywords}
galaxies: active -- galaxies: clusters: general -- X-rays: galaxies: clusters
\end{keywords}

%%%%%%%%%%%%%%%%%%%%%%%%%%%%%%%%%%%%%%%%%%%%%%%%%%

%%%%%%%%%%%%%%%%% BODY OF PAPER %%%%%%%%%%%%%%%%%%

\section{Introduction}

The environments of galaxies are expected to play a critical role in their evolution, influencing their morphology, star formation (SF) activity, and potentially regulating mass accretion onto their central supermassive black holes (SMBHs). Ram pressure can effectively strip the gaseous halos of satellite galaxies in groups and clusters \citep[e.g.][]{Gunn1972,Ebeling2014} and assist in guiding that gas to flow onto central galaxies, exacerbating gaseous wealth inequality amongst halo members. Dense environments can also ``strangle'' galaxies by preventing the accretion of ambient halo gas \citep[e.g.][]{Larson1980,Bekki2002}, halting the replenishment of the galaxy's gas supply. Galaxies can be tidally harassed by their neighbours and the cluster potential, which can deplete the cold gas reservoirs of satellites \citep[e.g.][]{Moore1996,Moore1999,Farouki1981} or perturb bound gas reservoirs, potentially triggering activity. Additionally, galaxies in dense environments are more likely to have undergone past major and minor mergers which can fundamentally change host-galaxy properties \citep[e.g.][]{Hopkins2006,Lin2010}. 

Multiwavelength studies of active galactic nuclei (AGN) accretion have identified two typical populations: a radiatively efficient population, characterized by a luminous accretion disk selected by X-ray and optical/UV studies, and a radiatively inefficient population, characterized by weak or absent X-ray/optical/UV emission and the presence of powerful relativistic jets, frequently observed through their radio emission. 
The radiatively efficient mode is thought to be driven by the accretion of cold gas onto the SMBH, which may drive the majority of SMBH mass growth. References to ``AGN'' in the rest of this paper will implicitly refer to the radiatively efficient AGN population.

Since both SF and AGN activity are thought to rely on steady supplies of cold gas fuel, environments that limit this fuel supply would be expected to suppress both phenomena. Furthermore, we would expect this effect to be most pronounced for satellites in massive galaxy clusters, where both the density of the intracluster medium (ICM) and number density of galaxies are the highest. This has indeed been established for SF in the local universe, where dense, cluster environments are more likely to host quiescent galaxies and less-dense, field regions host more star forming galaxies \citep[e.g.][]{Dressler1980,Kauffmann2004}. However, the situation for AGN activity is less clear. This is due not only to diversity in AGN accretion mode, degree of obscuration and host galaxy properties but also to differing AGN selection techniques and depths.

Early studies of the AGN population in clusters found that luminous AGN were less likely to be identified in clusters than the field \citep[e.g.][]{Gisler1978}. Several more recent studies with larger sample sizes have found similar results: Looking at quiescent galaxies in 521 SDSS clusters with $z<0.1$, \cite{vonderLinden2010} found that optical AGN activity is significantly suppressed within $\sim0.5R_{vir}$ (where $R_{vir}$ is the virial radius of the cluster) relative to the field. This environmental suppression of optical AGN activity was also found by \cite{Mo2018} in a sample of 2300 IR selected galaxy clusters at $z\sim1$. In a study of 32 galaxy clusters with $0.05<z<1.3$, \cite{Martini2009} found that the activity of X-ray selected, luminous AGN is suppressed within $\sim R_{vir}$ relative to the field. This is supported by \cite{Ehlert2014} who looked at the radial distribution of X-ray selected, luminous AGN in a sample of 42 clusters at $0.2<z<0.7$ and found a factor of three suppression of the AGN fraction within $\sim0.25R_{vir}$, but comparable values to the field at $R_{vir}$.

There is also a body of literature that has found no significant difference between AGN populations in galaxy clusters and the field. However, these studies have typically probed lower luminosity AGN. For instance, in a study of 33 galaxy clusters at $0.14<z<1.05$, \cite{Koulouridis2014} investigated low to moderate luminosity X-ray AGN activity within $0.5-2.5R_{vir}$ and found no evidence for environmental suppression. Similar findings were reported by \cite{Haggard2010} who found no significant difference between low-luminosity X-ray AGN populations in the field and in five clusters at $0.05<z<0.31$. Additionally, \cite{Pimbblet2013} found no difference between the optical AGN fraction within $1.5R_{vir}$ versus that at $2-3R_{vir}$ in a sample of six SDSS clusters at $z\sim0.07$.

When comparing AGN populations in high-density to low-density regions while explicitly controlling for absolute magnitude and/or stellar-mass, many studies have also found no significant differences \citep[e.g.][]{Powell2018,Yang2018}. However, because these studies rely on either contiguous or relatively small regions of sky, they have typically been unable to probe the most massive halos ($M>10^{15}M_\odot$) where any environmental influences would be most pronounced.

Furthermore, there is some evidence for a stronger evolution with redshift of the cluster AGN fraction than the field, such that at $z\gtrsim1$ enhancement of the luminous AGN population is observed \citep[e.g.][]{Martini2013,Krishnan2017}. This is most pronounced in observations of AGN activity in $z>2$ protoclusters \citep[e.g.][]{Lehmer2009,Digby-North2010,Lehmer2013,Umehata2015,Krishnan2017} although the enhancement is not unanimously observed \citep{Macuga2018}.

It is difficult to make direct comparisons between various studies in the literature as they often select environment and AGN in different ways. There is a great deal of diversity in the environments that are probed (i.e. filaments vs. groups vs. clusters vs. protoclusters) and even among studies solely looking at massive, virialized clusters, conclusions are often based on measurements made at different radii where different environmental interactions are pronounced. Furthermore, studies that select AGN of different luminosities and using varying wavelengths (i.e. optical vs. MIR vs. X-ray selection) are probing populations that may not be comparable.\\
\indent Optical AGN selection not only requires complete spectroscopy of a given parent population, but it is inherently biased towards luminous, unobscured sources. Furthermore, optical AGN diagnostics \citep{Kewley2006,Kauffmann2003} can confuse AGN and SFG at high redshifts \citep{Dickey2016}. MIR emission from a dusty AGN torus is often dwarfed by the host-galaxy emission due to dusty SF and these two contributions can be difficult to disentangle. This frequently leads to bona fide AGN being missed by MIR surveys and biases MIR AGN selection towards galaxies with less dust \citep[see][]{Hickox2009}. In contrast, X-ray observations allow us to robustly identify the radiatively efficient AGN population, directly probing emission from the immediate vicinity of the SMBH. X-ray emission can penetrate substantial hydrogen column densities and provides strong contrast versus host-galaxy starlight \citep[e.g.][]{Brandt2015,Xue2017}. However, while being the primary wavelength of choice for this study, X-ray selection has reduced sensitivity to compton-thick AGN \citep[e.g.][]{Li2019} and the rare population of intrinsically X-ray weak AGN \citep[e.g.][]{Teng2014,Luo2014}. \\
\indent This study is part of the Cluster AGN Topography Survey (CATS; Canning et al. in prep, King et al. in prep). CATS is a comprehensive, multi-wavelength survey of the AGN population in 487 galaxy clusters spanning redshifts $0.02<z<1.5$ and cluster masses $2\times10^{13}M_\odot<M_{500}<3.5\times10^{15}M_\odot$. It extends and expands on the methods developed in \cite{Ehlert2013,Ehlert2014,Ehlert2015}.
In this paper, we follow up the X-ray AGN population in 7 CATS clusters at $z\sim0.4$ with optical spectroscopy, allowing us to confirm cluster membership and compute the X-ray luminosities of our sources. In order to investigate the impact of large-scale environment on AGN activity, we apply identical selection criteria to the COSMOS control field and test for differences between the cluster and field populations. We additionally compare the inactive and active {\it cluster member} populations in our clusters to directly probe unique drivers of cluster AGN activity. In Section 2, we present our cluster sample along with our AGN selection methodology and data analysis techniques. Our results are presented in Section 3 and their implications are discussed in Section 4. \\
\indent All magnitudes quoted in this work are AB magnitudes. Distances are computed adopting a cosmology with $\Omega_{M}=0.3$, $\Omega_\Lambda = 0.7$, and $H_0=70~\mathrm{km~s^{-1}~Mpc^{-1}}$. Uncertainties are quoted at the 1$\sigma$ (68\%) confidence level. Cluster radii are measured in units of $r_{500}$, which is defined as the radius within which the mean density of the cluster is 500 times the critical density at that redshift. Cluster masses are quoted as $M_{500}$ values, with these being the mass contained within a sphere of radius $r_{500}$. They are inferred from the measured gas masses \citep{Mantz2016a}, which serve as a robust low-scatter mass proxies for the most massive systems \citep[e.g.][]{Allen2011}.

\section{Observations and data reduction}

\subsection{Cluster sample}

Our study includes 7 galaxy clusters drawn from the MAssive Cluster Survey \citep[MACS;][]{Ebeling2010} that are visible from the south with redshifts between $z=0.355$ and $z=0.451$ and $M_{500}$ values ranging from $5.5\times10^{14}~\mathrm{M_\odot}$ to $2.2\times10^{15}~\mathrm{M_\odot}$. These are some of the most massive and X-ray luminous clusters known, where we would expect any environmental influence on AGN activity to be most pronounced. We select a narrow redshift band in order to control for possible redshift evolution in the cluster AGN population. This particular band is well-suited to VIMOS spectroscopy (see \cref{sec:VIMOS}). All clusters have been observed with the Advanced CCD Imaging Spectrometer (ACIS) aboard the {\it Chandra} X-ray Observatory and have multi-band Subaru Suprime-Cam ($B_J,V_J,R_C,I_C,i^+,z^+$) photometry from the Weighing the Giants Survey \citep[WtG;][]{vonderLinden2014}. We determine the X-ray centres, masses, and radii of the clusters following the methodology of \cite{Mantz2016a}. All cluster properties are listed in \cref{tab:clusters}.

\begin{table*}
 \caption{Properties of the cluster sample used in this work. The RA and DEC are positions of the X-ray determined cluster centroids. All centroids, masses, and radii of the clusters are determined following the methodology of \protect\cite{Mantz2016a}. The {\it Chandra} exposure time only includes good time intervals. $\sigma_{cz}$ is the rest frame cluster velocity dispersion. $C_{X}$, $C_{S}$, and $C_{z}$ are the X-ray point source detection,  Subaru photometric, and VIMOS targeting completenesses respectively.
 }
 \label{tab:clusters}
 \begin{tabular}{lccccccccccccc}
  \hline
  Name &  $z$ &  RA  & Dec  & $M_{500}$ & $r_{500}$ & {\it Chandra} exp. & $\sigma_{cz}$ & $C_X$ & $C_S$ & $C_z$ \\
 &  & (deg) & (deg) & ($10^{15}~\mathrm{M_\odot}$) & (kpc) &  (ks)  & (km/s) & & \\
  \hline
 
  MACSJ1115.8$+$0129 & 0.355 & 168.96606 & 1.49898 & $0.81\pm0.14$     & $1250\pm70$ & 44.3 & 793            &  0.85 & 0.79  & 0.33     \\[0.1cm]
  MACSJ2211.7$-$0349 &  0.397 & 332.94129 & $-$3.83006 & $1.8\pm0.3$   & $1590\pm90$ &13.4 & 954             &  0.69 & 0.94  & 0.33   \\[0.1cm]
  MACSJ0429.6$-$0253 & 0.399 & 67.40004 & $-$2.88526 & $0.55\pm0.11$   & $1080\pm70$ & 19.3 & 899           & 0.91  & 0.80  & 0.50   \\[0.1cm]
  MACSJ0451.9$+$0006 &  0.429 & 72.97725 & 0.10579 & $0.77\pm0.17$  &  $1190\pm90$ & 9.7 & 855                &   0.90 & 0.73 & 0.90  \\[0.1cm]
  MACSJ0417.5$-$1154 &  0.443 & 64.39453 & $-$11.90916 & $2.2\pm0.3$  & $1700\pm90$ & 81.5 & 1077           &  0.86  & 0.87 & 0.79  \\[0.1cm]
  MACSJ0329.6$-$0211 & 0.450 & 52.42343 & $-$2.19650 & $0.70\pm0.13$     & $1150\pm70$ & 22.2 & 801         & 0.96  & 0.90  & 1.00  \\[0.1cm]
  MACSJ1347.5$-$1144 &  0.451 & 206.87768 & $-$11.75239 & $1.7\pm0.3$ & $1540\pm90$ &  206.5 & 824         &  0.84 & 0.92  & 1.00  \\
  \hline
 \end{tabular}
\end{table*}

\begin{figure}
\begin{center}
\includegraphics[width=0.5\textwidth]{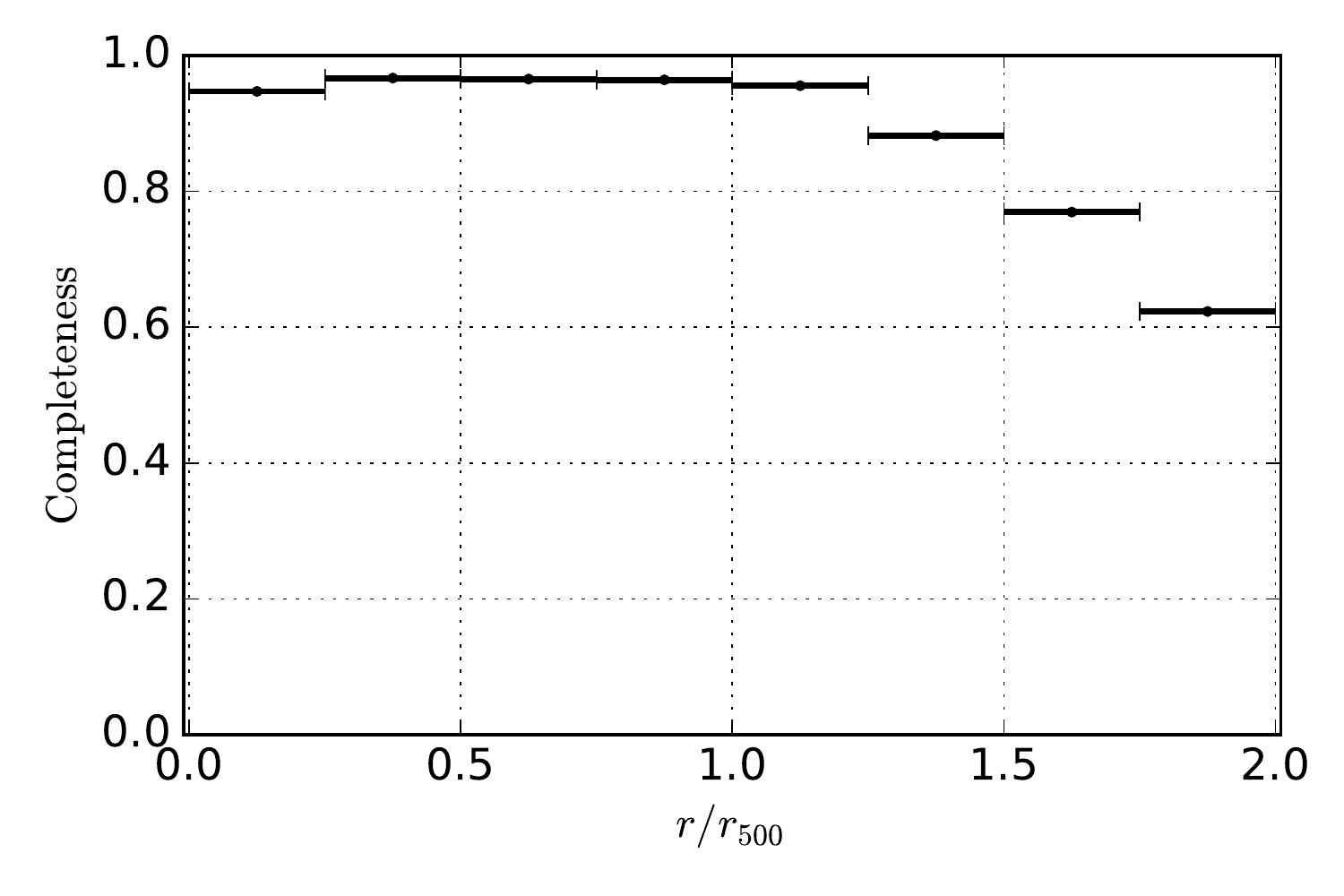}
\caption{The completeness of our X-ray point source detection for objects with $F_X(0.5-8~\mathrm{keV})>10^{-14}~\mathrm{erg~cm^{-2}~s^{-1}}$ as a function of cluster-centric radius. The completeness is $\sim96\%$ within $r_{500}$ but drops slightly near the cluster core due to the presence of diffuse cluster emission, and significantly at larger radii due to both a worsening of the {\it Chandra} PSF and incomplete {\it Chandra} coverage of some clusters. This incompleteness is corrected for in the analysis.}  \label{fig:xagncomplete}
\end{center}
\end{figure}

\begin{figure*}
\begin{center}
\includegraphics[width=0.8\textwidth]{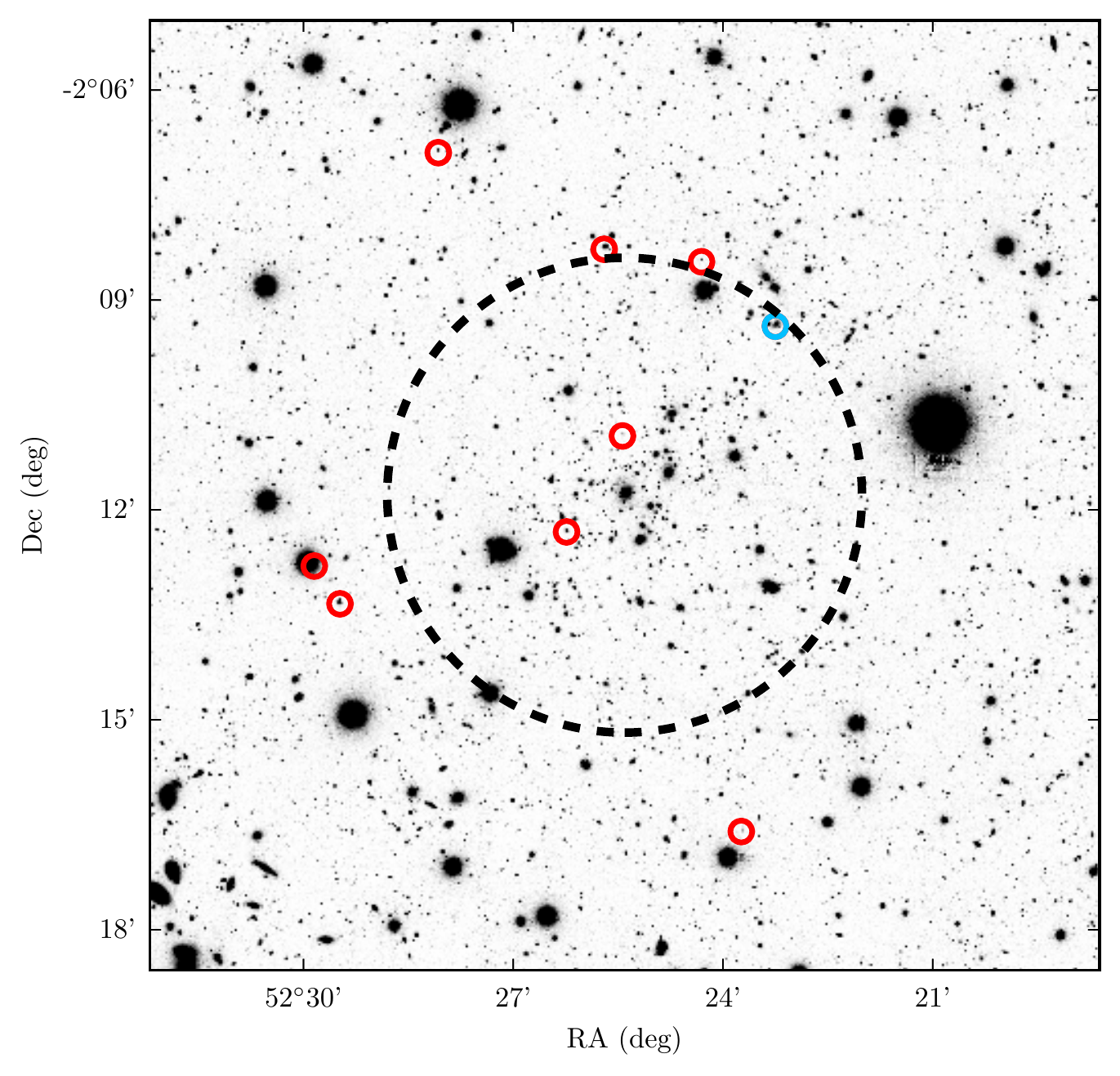}
\caption{Subaru Suprime-Cam $V$-band image of MACS0329.6-0211. All X-ray point sources above our luminosity threshold are marked by red circles. The single AGN that is spectroscopically identified with the cluster is circled in blue. The dashed black circle outlines $r_{500}$ and is centred on the cluster X-ray centroid.
}  \label{fig:cluster_image}
\end{center}
\end{figure*}

\begin{figure} 
\begin{center}
\includegraphics[width=0.5\textwidth]{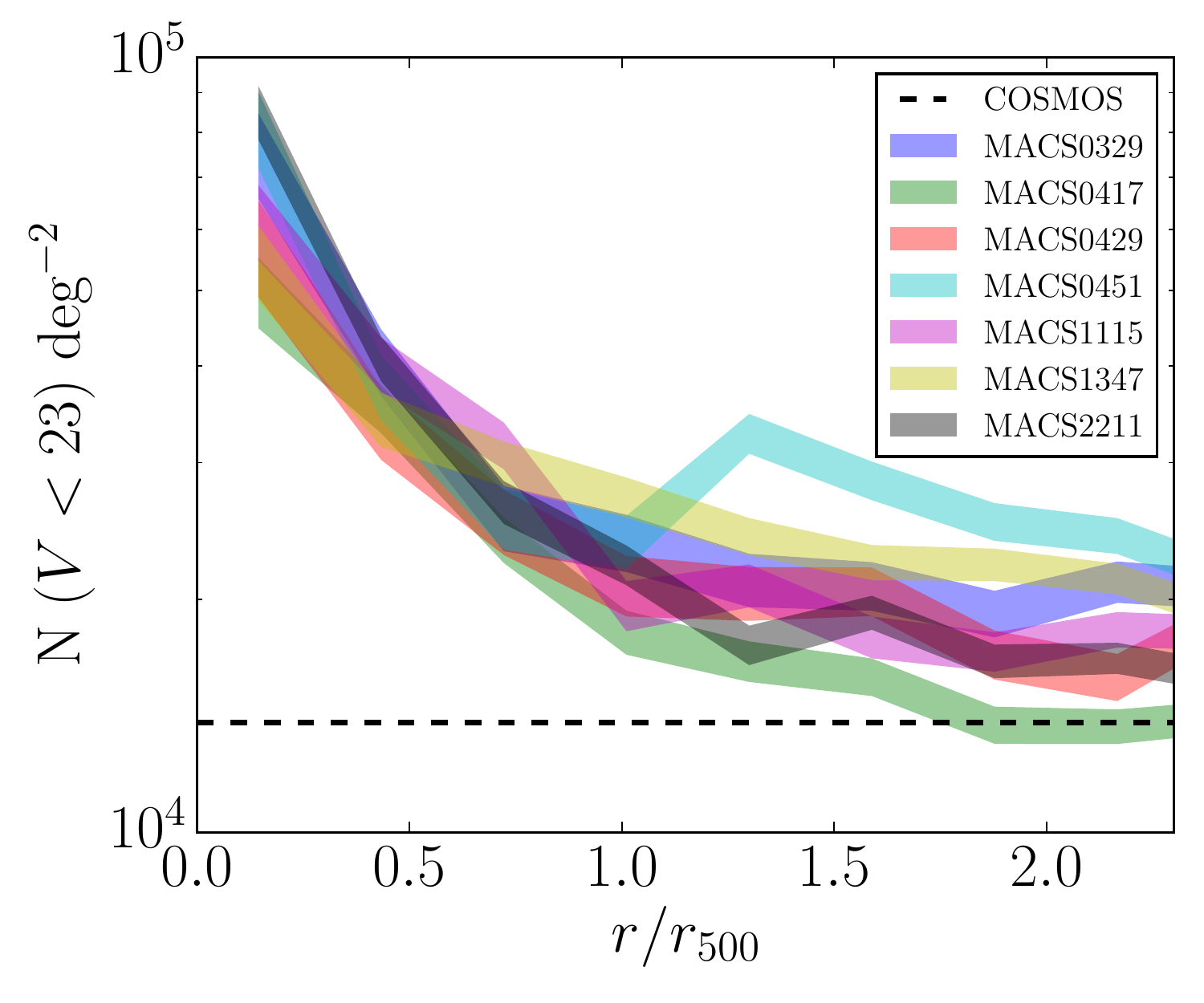}
\caption{Shaded regions show the 1$\sigma$ contours of the projected number density of galaxies with $V<23$ as a function of cluster-centric distance for each of the clusters in our sample. The COSMOS field density is shown in dashed black and is subtracted from the projected number density of each cluster to recover the surface density of cluster member galaxies.
}\label{fig:galdensity}
\end{center}
\end{figure}

\subsection{X-ray point source detection} \label{sec:xray}

Our clusters have between 10 and 207 ks of clean, archival {\it Chandra} exposures which we use to identify X-ray point sources in the cluster fields. Our X-ray point source detection technique is described in detail in Canning et al. in prep. In short, we first run \verb+WAVDETECT+ \citep{Freeman2002} optimized to maximize the completeness of our catalog. We then follow up each source using the \verb+ACIS-EXTRACT+ code \citep{Broos2010} to filter out extended and spurious sources, maximizing the purity of our sample. For sources on top of diffuse cluster emission, a local background is modelled and the source profile is compared to the {\it Chandra} PSF at the source location. For the present study, we limit our X-ray point source catalogs to sources that fall within $2r_{500}$ (approximately the virial radius) of the X-ray determined cluster center. 

We initially impose a 0.5-8 keV flux limit of $>10^{-14}~\mathrm{erg~cm^{-2}~s^{-1}}$ on our sample of X-ray point sources. The average completeness at this flux limit is 85\% but there is a significant radial dependence as shown in \cref{fig:xagncomplete}. While we are $\sim96\%$ complete within $r_{500}$, our completeness falls slightly near the cluster core due to the presence of relatively bright, diffuse cluster emission; and significantly at larger radii due to both a worsening of the {\it Chandra} PSF and incomplete {\it Chandra} coverage of some clusters (see Canning et al. in prep, for details). Our overall X-ray point source detection completeness $C_X$ for each cluster is listed in \cref{tab:clusters}. Incompleteness is accounted for as detailed in \cref{sec:massdep}.

We find a total of 165 X-ray point sources above this flux limit.  All of these sources have a no-source binomial probability $<10^{-4}$ \citep[see appendix A of][]{Weisskopf2007} implying a negligible number of false source detections. 

\subsection{Subaru photometry} \label{sec:subaru}

We use deep Subaru Suprime-Cam imaging from WtG to provide accurate relative astrometry for the X-ray AGN and to identify the general galaxy population in the clusters. A sample $V$-band image of MACS0329.6-0211 is shown in \cref{fig:cluster_image}. We utilize the optical catalogs described in Section 6.2 of \cite{vonderLinden2014}. These catalogs were built using SExtractor \citep{Bertin1996} with settings optimized to identify extended objects and are particularly suited for the photometry relevant to this study. The catalogs are 73-94\% complete at $r\leq2r_{500}$ and $V_J<23$ and have accurate astrometry to better than $0.1''$. The Subaru photometric incompleteness $C_S$ for each cluster is listed in \cref{tab:clusters}. Note that the optical incompleteness is largely due to the masking of saturated objects and artefacts around them. Incompleteness is accounted for in our analysis on a cluster-by-cluster basis as detailed in \cref{sec:massdep}.
We use $3''$ aperture magnitudes to match the Suprime-Cam photometry of the COSMOS field, which we use as our field control sample as outlined in  \cref{sec:field}. We distinguish between stars and extended objects using the full width at half-maximum and the SExtractor \verb+CLASS_STAR+ parameter \citep[see][for details]{vonderLinden2014}.

We visually identify the Brightest Cluster Galaxy (BCG) in each cluster using the Subaru imaging and exclude them from this study as they may have formed by/follow different physical processes than the general cluster galaxy population. However, only the BCG of MACSJ1347.5$-$1144 is found to be X-ray luminous with a 0.5-8 keV luminosity of $7.6\times10^{43}~\mathrm{erg~s^{-1}}$. We then construct the projected galaxy number densities for each cluster as a function of cluster-centric distance, as shown in \cref{fig:galdensity}. We compute the expected field density of galaxies with the same $V$-band magnitude cut from COSMOS \citep{Laigle2016} and subtract this from the projected number density in order to recover the number density of cluster member galaxies. 

\subsection{Counterpart matching} \label{sec:counter}

To generate the parent list which we target for spectroscopy, we first match our X-ray and optical photometric catalogs. For each source $i$ in the X-ray catalog, we search for optical matches in the Subaru catalog based on source-by-source X-ray centroiding uncertainties, $\sigma_{X,i}$. These uncertainties are determined by simulating point sources through our detection pipeline as detailed in Canning et al. in prep. For each object $i$, we identify optical counterparts within a projected distance $D_i=\sqrt{(3\sigma_{X,i})^2+(0.5'')^2}$ where the latter term is to account for uncertainty in the {\it Chandra} astrometry. In 7 cases, more than one optical counterpart was identified and the nearest match was taken as the counterpart. For objects spectroscopically identified as cluster members (see \cref{sec:member}), counterparts were unambiguous. Uncertainties in our optical centroiding and astrometry are negligible relative to our X-ray positional errors and are not taken into account when determining counterparts.

We remove all X-ray point sources that are matched to heavily saturated objects in the optical as they are typically associated with bright stars. On average there are 2 of these objects per cluster with $F_X>10^{-14}~\mathrm{ergs~cm^{-2}~s^{-1}}$. We further limit our study to sources with optical counterparts brighter than $V=23$ where we can reliably determine redshifts (see \cref{sec:spec+redshift}). This defines our parent sample of 56 X-ray point sources with $F_X>10^{-14}~\mathrm{erg~cm^{-2}~s^{-1}}$, $r<2r_{500}$, and $V<23$.

\subsection{VLT spectroscopy} \label{sec:VIMOS}

We obtained spectroscopy of the cluster fields using the VIsible Multi-Object Spectrograph (VIMOS) instrument in medium resolution (MR) mode on the Very Large Telescope (VLT). The MR mode covers the wavelength range 480-1000 nm with a spectral resolution of 580. Through ESO programmes 090.A-0958(B), 092.A-0405(A), and 094.A-0557(A), we successfully observed every cluster with $1-4$ multi-object masks, each with an average $\sim$2700 second exposure.
Every object in the parent sample was assigned a compulsory flag in the VIMOS Mask Preparation Software (VMMPS) and had a high chance of being targeted. We observed each cluster with several masks, dithered across the cluster to maximize our coverage.
Each mask is observed several times
giving us multiple independent spectral measurements for many objects.
Across all clusters, our average spectral targeting completeness of the parent X-ray population is 60\% due to slit packing limitations and weather based observing failures. However, our successful targets are effectively randomly selected, independent of optical magnitude, X-ray flux, and cluster-centric radius. The VIMOS targeting completeness $C_z$ for each cluster is listed in \cref{tab:clusters}. This incompleteness is accounted for in the analysis below on a cluster by cluster basis, as detailed in \cref{sec:massdep}.

We fill the rest of our MOS masks with the general galaxy population with $V<23$. This allows us to not only measure the cluster velocity dispersion, which is used when determining cluster membership (see \cref{sec:member}), but also provides a sample of inactive member galaxies, to which we compare to our AGN hosts in \cref{sec:density,sec:phase_space}.

\subsection{Spectral reduction and redshift determination} \label{sec:spec+redshift}

\begin{figure*} 
\begin{center}
\includegraphics[width=\textwidth]{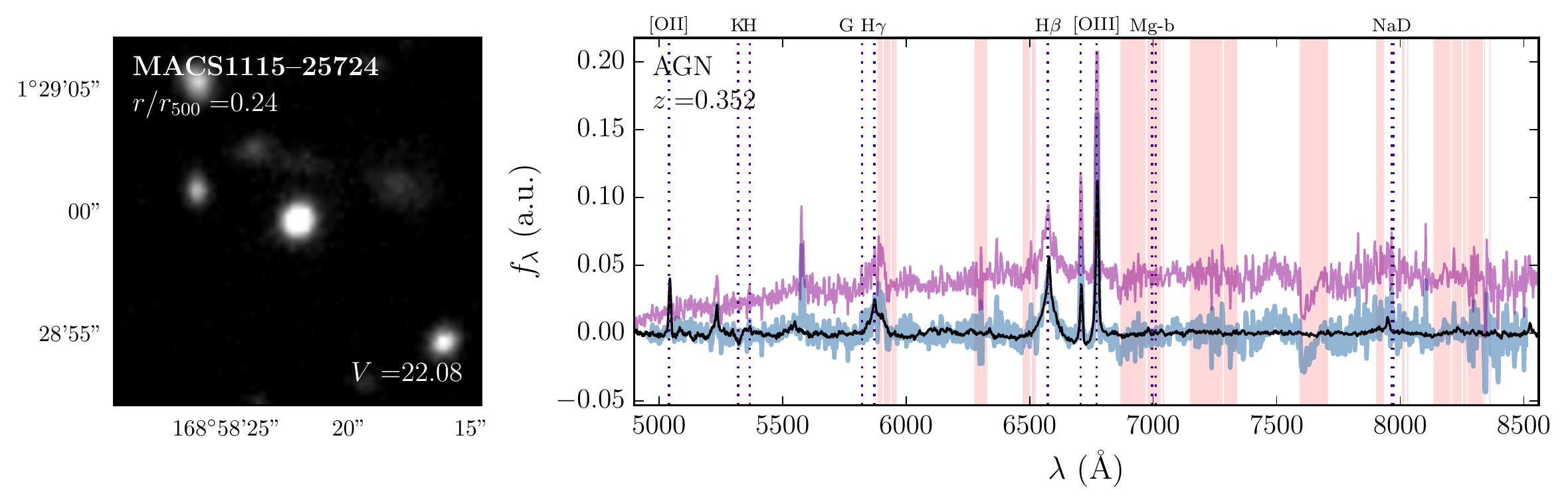}
\vspace{-1em}
\caption{{\bf Left:} $V$-band Subaru imaging $15''\times15''$ cutout of a sample X-ray point source (centered in the image) that is a confirmed cluster member. {\bf Right:} VIMOS 1D extracted spectrum is shown in purple, the continuum subtracted spectrum in blue, and the best fitting PCA reconstruction shown in black. Telluric absorption bands that were masked in the redshift fitting procedure are shaded red. Imaging and spectra for all seven confirmed cluster member AGN can be found in \cref{app:spectra}.} \label{fig:spectrum}
\end{center}
\end{figure*}

We use the EsoReflex automated data reduction workflow for VIMOS spectroscopy \citep{esoreflex} to extract 1D spectra from our 2D Multi-Object Spectra (MOS). 
Multiple independent exposures of the same object are stacked to maximize the signal-to-noise ratio. 

We use code based on the SDSS \verb+idlspec2d+ \citep{Bolton2012} and DEEP2 software \citep{Newman2013} to perform the spectral fitting and subsequent redshift determination of our objects. Classification templates for galaxies, quasars, Cataclysmic Variable (CV) stars, and non-CV stars are built from large samples of the respective objects, with known redshifts, corrected to their rest frame. Principal Component Analysis (PCA) is performed on these sets of spectra and the leading ``eigenspectra" compose the basis from which linear combinations are made to create a model spectrum for a given object \citep[following][]{Glazebrook1998}. 

For each spectrum, we first perform continuum subtraction by subtracting off the median of a 480 \AA~wide moving boxcar in each wavelength bin. This is also performed for all eigenspectra following \cite{Glazebrook1998}. We then perform a coarse iteration over trial redshifts and at each redshift compute the $\chi^2$ value for each of the best fitting galaxy, quasar, CV, and non-CV star models. 
The five best fitting redshifts for each class are then re-visited and finely resampled at the sub-pixel level to arrive at a final list of redshifts and associated $\chi^2$ values for each class. The redshift and class that yield the best fit to the input spectrum is adopted as our final measurement. We refer the reader to \cite{Bolton2012} for further details on the algorithm. A sample reduced 1D spectrum and the associated best-fitting linear combination of eigenspectra is shown in \cref{fig:spectrum}.

The pipeline outlined above computes the statistical redshift uncertainty directly during the $\chi^2$ minimization routine
but catastrophic misclassifications are left unidentified. We visually inspect the spectral fits for all X-ray point sources and remove any obvious catastrophic failures (nine were so identified). None of the sources that were removed in this procedure were initially assigned a redshift consistent with a cluster. We expect an insignificant redshift failure rate for $V<23$ objects that are within the $0.35\lesssim z\lesssim0.45$ window for which our VIMOS program was designed. A typical quiescent galaxy with $V=23$ has a spectral signal-to-noise ratio of $S/N>5$ per pixel in our VIMOS setup, which is enough to determine a redshift reliably. Visual inspection of spectral fits confirms our negligible redshift failure rate for such sources. A catalog of the 1912 reliable redshifts acquired in this study are made available online as supplementary material.

\begin{table*}
 \caption{Properties of the identified cluster member AGN.
 }
 \label{tab:agn}
 \begin{tabular}{lcccccccc}
  \hline
  Host cluster & RA  & Dec  & $r/r_{500}$ & $L_X$(0.5-8 keV) & $V$ & Net counts & HR  & Optical classification \\
  & (deg) & (deg) & & ($10^{42}~\mathrm{ergs~s^{-1}}$) & (mag) & (0.5-8 keV) &    & \\
  \hline

  MACSJ0329.6$-$0211 & 52.38739 & -2.15626 & 0.95 & 12.6 & 20.81 & 31 & 0.77 & Inactive \\
  MACSJ0451.9$+$0006 &  72.95817 & 0.11557 & 0.35 & 9.42 & 21.49 & 13 & -0.70 & Inactive \\
  MACSJ0451.9$+$0006 & 72.98720 & 0.10894 & 0.17 & 6.95 & 21.73 & 10 & -0.20 &  Type II \\
  MACSJ0451.9$+$0006 &  73.09899 & 0.09693 & 2.00 & 9.14 & 21.04 & 9 & 0.10 & Type II \\
  MACSJ0451.9$+$0006 &  73.05900 & 0.03219 & 1.80 & 15.6 & 21.61 & 19 & -0.38 & Inactive \\
  MACSJ1115.8$+$0129 &  168.97280 & 1.48328 & 0.24 & 9.50  & 22.08 & 91 & -0.33 & Type I \\
%  MACSJ1347.5$-$1144 & 206.87746 & -11.75255 & 0.00 & 75.5 & 19.96 & 2112 & -0.38 BCG \\
 MACSJ1347.5$-$1144 & 206.86774 & -11.87376 & 1.61 & 12.9 & 21.92 & 138 & -0.33 & Type II \\
  \hline
 \end{tabular}
\end{table*}

\subsection{AGN identification}

Assuming an X-ray power-law photon index of $\Gamma=1.7$ (our results are qualitatively insensitive to a choice of $1.5<\Gamma<1.9$), our flux limit corresponds to a rest-frame 0.5-8 keV luminosity of $6.8\times10^{42}~\mathrm{erg~s^{-1}}$ at the redshift of our furthest cluster, $z=0.451$. Above this threshold, we expect negligible contamination from star forming galaxies as they are typically characterized by 0.5-8 keV luminosities below $3\times10^{42}~\mathrm{erg~s^{-1}}$ \citep[e.g.][]{Bauer2004}. We constrain our sample of AGN to objects above this luminosity threshold. We find a total of 49 X-ray AGN after making this cut.

\subsection{Cluster membership} \label{sec:member}

To determine the cluster membership of our AGN we use our full spectroscopic sample of cluster galaxies to measure the cluster velocity dispersion. This is done by first cutting out all objects that fall outside $\delta z=0.02$ of the cluster redshift and then iteratively sigma-clipping $3\sigma$ outliers from the remaining distribution. There are an average of 44 spectroscopic members remaining per cluster after this procedure. The cluster velocity dispersion ($\sigma_{cz}$) is computed as the standard deviation in the redshifts of the remaining objects. These are corrected to the cluster rest frame and listed in \cref{tab:clusters}. From this sample we also determine the mean recession velocity of the cluster $\overline{cz}$ which in all cases is in good agreement with the literature redshifts given in \cref{tab:clusters}. Cluster membership is then defined as all galaxies that fall within $\pm3\sigma_{cz}$ of $\overline{cz}$. We expect the large majority of these galaxies to be associated with the virialzed cluster volume.

From our sample of 49 AGN in the seven cluster fields, we find that seven of them are genuine cluster members. The details of these AGN are outlined in \cref{tab:agn}. A spectrum of one of these cluster member AGN is shown in \cref{fig:spectrum} alongside the best fitting PCA reconstruction and $V$-band Subaru imaging. 

\subsection{Control field} \label{sec:field}

In order to test whether the cluster AGN population differs from that of the field, we need a reliable control sample. For this we utilize the COSMOS field. Starting from the COSMOS catalog of \cite{Laigle2016} we restrict our analysis to the 1.38 deg$^2$ UltraVISTA area inside the COSMOS 2deg$^2$ field, after removing regions with bad and saturated pixels in the optical and NIR. We further filter the catalog to objects with $V<23$ to match our cluster analysis, well above the $3\sigma$ depth of COSMOS. The X-ray measurements of the field population come from the {\it Chandra} COSMOS Legacy Survey \citep{Civano2016} and we utilize photometric redshifts from \cite{Marchesi2016}.
To match our cluster sample, we restrict the field to $0.35<z<0.45$ and we constrain the X-ray population to sources with $0.5-8$~keV luminosities above our threshold of $6.8\times10^{42}~\mathrm{erg~s^{-1}}$. There are nine X-ray AGN in the COSMOS field that satisfy these criteria.

\begin{table}
 \caption{Cluster AGN properties. $N_{AGN}$ is the number of cluster member AGN identified within $2r_{500}$ of each cluster. $f_{AGN}$ is the completeness corrected fraction of $r<2r_{500}$ cluster member galaxies that are host to an AGN listed with $1\sigma$ binomial uncertainties. $\Phi$ is the completeness adjusted space density of cluster member AGN quoted with $1\sigma$ Poisson uncertainties.}
 \label{tab:cluster_agn}
 \begin{tabular}{lccc}
  \hline
  Cluster &  $N_{AGN}$ & $f_{AGN}$ & $\Phi$  \\
 & & \% & ($10^{-2}$ Mpc$^{-3}$)  \\
  \hline
 
  MACSJ1115.8$+$0129 & 1 &    $0.6^{+1.4}_{-0.2}$  &  $6.7^{+15.7}_{-5.6}$    \\[0.1cm]
  MACSJ2211.7$-$0349 &  0 &   $0^{+2.0}$                &   $0^{+6.2}$  \\[0.1cm]
  MACSJ0429.6$-$0253 & 0 &   $0^{+2.1}$                 &    $0^{+11.9}$ \\[0.1cm]
  MACSJ0451.9$+$0006 &  4 & $0.5^{+0.4}_{-0.2}$    &    $11.6^{+9.2}_{-5.5}$  \\[0.1cm]
  MACSJ0417.5$-$1154 &  0 &   $0^{+0.5}$                &    $0^{+1.9}$  \\[0.1cm]
  MACSJ0329.6$-$0211 & 1 &    $0.3^{+0.8}_{-0.1}$   &  $2.2^{+5.2}_{-1.8}$     \\[0.1cm]
  MACSJ1347.5$-$1144 &  1 &   $0.2^{+0.4}_{-0.1}$   &   $1.0^{+2.4}_{-0.9}$   \\
  \hline
 \end{tabular}
\end{table}

\section{Results}

In the results below, the cluster AGN population refers to sources with $L_X>6.8\times10^{42}~\mathrm{erg~s^{-1}}$, $V<23$, and $r<2r_{500}$ that are cluster members. The inactive cluster galaxy population refers to all other members with $V<23$ and $r<2r_{500}$. Incompleteness due to X-ray point source detection (\cref{sec:xray}), Subaru photometry (\cref{sec:subaru}), and VIMOS targeting (\cref{sec:spec+redshift}) is corrected for on a cluster by cluster basis by adjusting the cluster effective volumes and projected areas as appropriate. The same X-ray luminosity and $V$-band magnitude cuts applied to the cluster sample are applied to the field. We also apply a redshift restriction of $0.35<z<0.45$ to the field data, matching the redshift range of the clusters in this study.

\subsection{AGN space density and cluster mass} \label{sec:massdep}

For each cluster, we compute the AGN space density $\Phi$ as
\begin{align}\label{eq:Phi}
\Phi=\frac{N}{V_{eff}}
\end{align}
where $N$ is the number of cluster member AGN identified and $V_{eff}$ is the effective proper volume of the cluster out to $2r_{500}$ and taking into account our completeness. This spherical volume is computed as
\begin{align}\label{eq:Veff}
V_{eff}=\frac{4}{3}\pi (2r_{500})^3 C_XC_SC_z
\end{align}
where $C_{X}$, $C_{S}$, and $C_{z}$ are the X-ray point source detection,  Subaru photometric, and VIMOS targeting completenesses respectively. The space densities of AGN in each cluster are tabulated along with their $1\sigma$ Poisson uncertainties in \cref{tab:cluster_agn}.

Virial arguments show that the galaxy velocity dispersions $\sigma$ in clusters scale with cluster mass as $\sim M^{1/3}$ and theoretical calculations suggest that the rate of galaxy mergers between cluster galaxies scales as $\sim\sigma^{-3}$ \citep{Mamon1992}. Thus, if galaxy-galaxy mergers were driving AGN activity in clusters, we would expect the AGN space density to scale as $\sim M^{-1}$. 

This motivates us to model the space density as a power law in cluster mass, such that
\begin{align}\label{eq:Phi^}
\widehat{\Phi}=\beta \left( \frac{M_{500}}{10^{15}\mathrm{M_\odot}}\right)^\alpha.
\end{align}

Given that the AGN counts in each cluster are Poisson distributed, the log likelihood of our data is
\begin{align}\label{eq:logL}
\ln L=\sum_{i=1}^{7} \ln \left[ \frac{e^{-\lambda}\lambda^{N_{i}}}{N_{i}!} \right] .
\end{align}

Here $N_i$ is the number of AGN observed in cluster $i$ and $\lambda=\widehat{\Phi} V_{eff}$ is the number of AGN we would expect to observe given our model of the space density, $\widehat{\Phi}$, and the effective cluster volume probed $V_{eff}$. Note that the $\sim15\%$ uncertainties on the cluster masses \citep{Applegate2014,Mantz2016a} are ignored in our modelling procedure as they are negligible relative to the large Poisson uncertainties associated with the small number of AGN observed in each cluster. This is confirmed through Monte Carlo simulation.

\begin{figure}
\begin{center}
\includegraphics[width=0.48\textwidth]{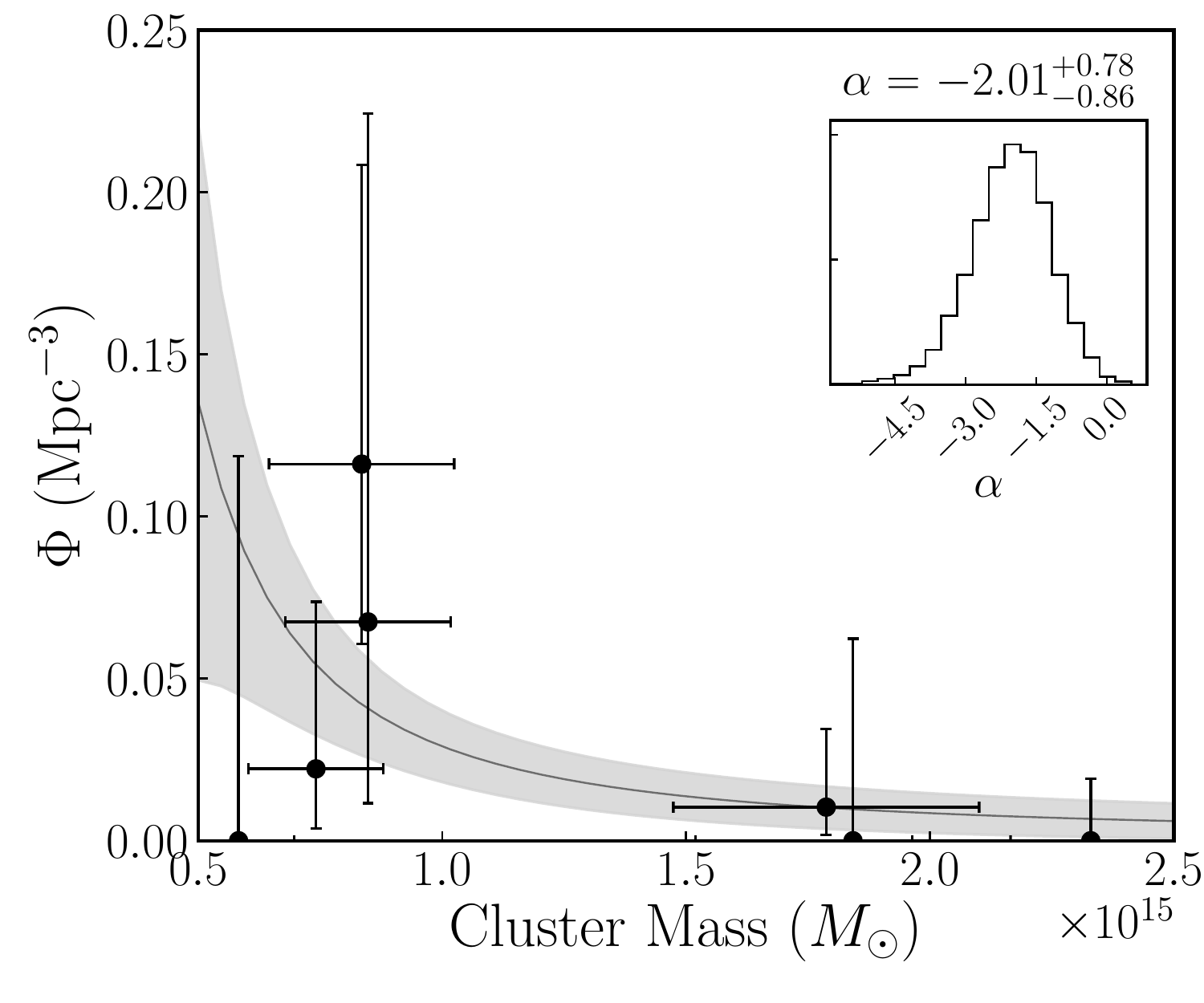} 
\vspace{-1.5em}
\caption{The space density $\Phi$ of cluster member X-ray AGN as a function of cluster mass is shown as black points with $1\sigma$ Poisson error bars. The space density is modelled as a power law in cluster mass where the AGN counts are Poisson distributed with an expected space density $\widehat{\Phi}\propto M^\alpha$. The black solid curve shows the best fitting model with the grey band highlighting the 1$\sigma$ uncertainty on the fit. The inset shows the PDF of the mass dependence as described in \cref{sec:massdep}.
}  \label{fig:nden_cm}
\end{center}
\end{figure}

We fit for $\alpha$ using the Affine Invariant Markov Chain Monte Carlo (MCMC) package \verb+emcee+ \citep{emcee}. The cluster AGN space density is plotted as a function of cluster mass in \cref{fig:nden_cm}. The inset on the left of \cref{fig:nden_cm} shows the posterior distribution for the power law dependence of the expected AGN space density on cluster mass, $\alpha=-2.0^{+0.8}_{-0.9}$. Our results rule out zero mass dependence at the $2.5\sigma$ level. The error bars plotted for illustration in \cref{fig:nden_cm} and quoted in \cref{tab:cluster_agn} are derived from the Gehrels $1\sigma$ Poisson uncertainties for small numbers of events \citep{Gehrels1986}. Note that these errors are not used in our MCMC fitting procedure outlined above.

We also perform the same model fit while excluding MACSJ0451.9$+$0006, as it may contain a uniquely enhanced AGN population (see \cref{sec:macs0451}). The resulting power law dependence is found to be $\alpha=-1.6^{+1.0}_{-1.1}$.

\subsection{Local galaxy density} \label{sec:density}

\begin{figure*}
\begin{center}
\includegraphics[width=0.49\textwidth]{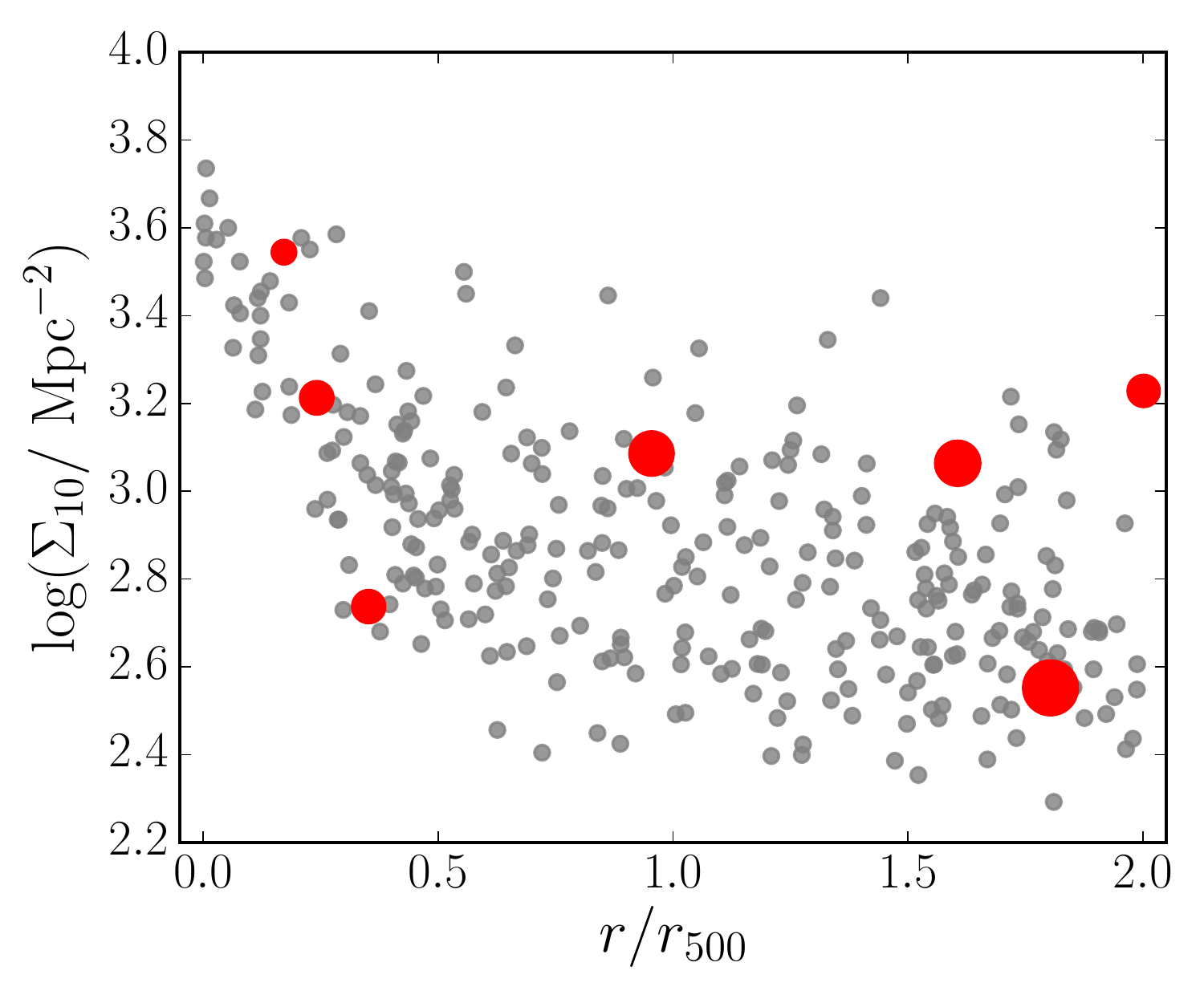}
\includegraphics[width=0.49\textwidth]{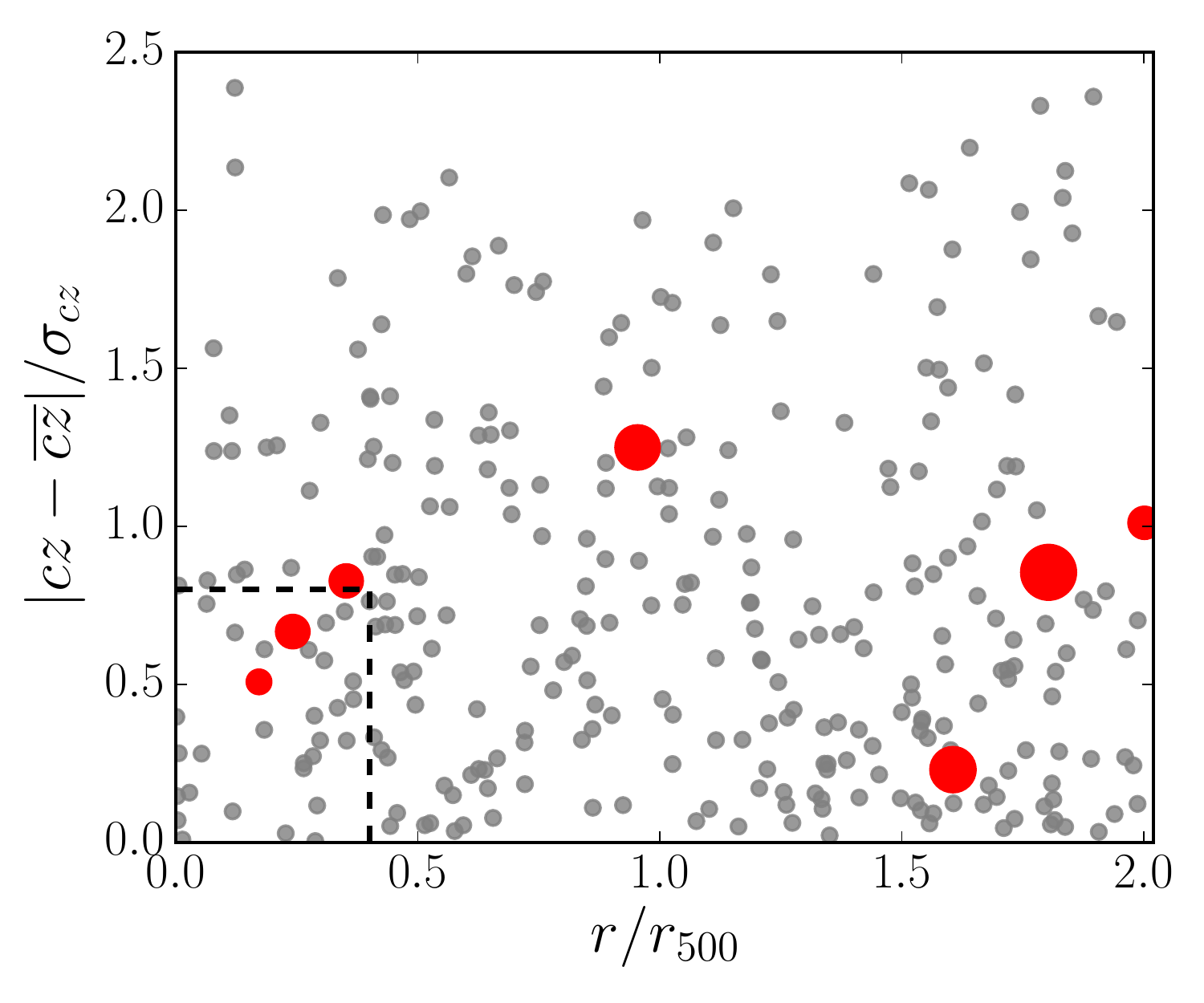}
\caption{{\bf Left:} The projected local density is plotted as a function of cluster centric distance for galaxies with $V<23$. Inactive cluster member galaxies are marked in grey while the cluster X-ray AGN population is marked by red circles with their size proportional to their X-ray luminosity. There is no statistically significant difference between the distributions of the two populations. {\bf Right:} The $|cz-\overline{cz}|/\sigma_{cz}$ vs. $r/r_{500}$ phase-space of our cluster member population is shown. The symbols have the same meanings as in the left panel. The region devoid of any AGN in the analysis of \protect\cite{Haines2012} is outlined in dashed black and contains two AGN in our sample. There is no statistically significant difference between the distributions of the two populations and we find no evidence that AGN preferentially lie along infalling caustics with $|cz-\overline{cz}|/\sigma_{cz}\gtrsim1$.
%{\CR (Plot inactive as contours?)}
} \label{fig:local_density}
\end{center}
\end{figure*}

To test whether AGN preferentially lie in over-dense regions within the clusters (for example, within merging sub-clusters), we compute the projected local galaxy density $\Sigma_{10}=10/A$ where $A$ is the projected circular area on the sky that encloses the 10 nearest galaxies in projection with $V<23$. $\Sigma_{10}$ is computed for every spectroscopically confirmed cluster member, both active and inactive, and plotted in the left panel of \cref{fig:local_density}. The AGN are plotted in red with their symbol size proportional to their X-ray luminosity. The radii of circles enclosing the 10 nearest neighbors range from $\sim25$ kpc near the cluster core to $\sim100$ kpc at the virial radius. We use the Fasano and Franceschini variant of the Peacock test \citep{Peacock1983,Fasano1987,Press2007} to look for any difference between the distributions of the active and inactive galaxies and find $P=0.17$, suggesting no statistically significant difference between the two populations.\footnote{The Peacock test is a generalization of the Kolmogorov-Smirnov test suitable for comparing two-dimensional distributions.} This result is independent of the $V$-band magnitude cut applied and also holds for the $\Sigma_{5}$ measure of local galaxy density.

\subsection{AGN in phase-space} \label{sec:phase_space}

We plot the $|cz-\overline{cz}|/\sigma_{cz}$ vs. $r/r_{500}$ phase-space of our cluster population in the right panel of \cref{fig:local_density} with AGN shown as red circles with size proportional to their X-ray luminosity. We again use the Fasano and Franceschini variant of the Peacock test \citep{Peacock1983,Fasano1987,Press2007} to look for any difference between the distributions of the active and inactive galaxies and find $P=0.70$, suggesting no statistically significant difference between the two populations. 

\subsection{Cluster AGN fraction vs. field}

For each cluster, we compute the fraction of member galaxies hosting an X-ray AGN by dividing the cluster AGN surface density $n=N/(\pi(2r_{500})^2C_XC_SC_z)$ by the cluster member surface density (see \cref{sec:subaru}). These values are tabulated along with their $1\sigma$ binomial uncertainties in \cref{tab:cluster_agn}. We perform the same calculation for the COSMOS field control sample.
Both the cluster and field AGN fractions are plotted as a function of $V$-band magnitude limit ($V_{max}$) in \cref{fig:fagn_vmax}. All cluster fractions are for $r<2r_{500}$ and are corrected for incompleteness both due to X-ray point source detection (see \cref{sec:xray}) and due to VIMOS targeting (see \cref{sec:spec+redshift}). While the cluster and field are consistent at faint magnitude cuts ($V_{max}>22$), the cluster AGN fraction is suppressed relative to the field in the brightest galaxies. Since we are selecting a narrow redshift slice $0.35<z<0.45$, this effectively corresponds to a suppression in the AGN fraction in clusters for the most massive galaxies. Furthermore, while the field AGN fraction increases monotonically with brightness \citep[due to a strong stellar mass dependence of field AGN activity, e.g.][]{Xue2010,Yang2018}, the cluster AGN fraction appears to have no dependence on the host galaxy magnitude.

\section{Discussion} \label{sec:disc}

The CATS survey aims to answer the questions: 1) does AGN activity depend on environment and, if so, 2) what drives this dependence. 
Our observations of seven massive galaxy clusters with $0.35<z<0.45$ have identified a total of seven X-ray AGN as being bona fide, intrinsic cluster members. We have compared their properties to those of their host cluster members and to an identically selected sample of field AGN from the COSMOS survey. Below we discuss the implications of our results on the two questions posited above.

\subsection{Does AGN activity depend on environment?} \label{sec:nden_disc}

The most significant result in this study is the inverse dependence of the cluster AGN space density on cluster mass, which scales as $\sim M^{-2.0^{+0.8}_{-0.9}}$ (see \cref{fig:nden_cm}) suggesting, at the $2.5\sigma$ level, that AGN activity does indeed depend on environment. This result is in agreement with the $\sim M^{-1.2\pm0.7}$ scaling relation found by \cite{Ehlert2015} in a photometric study of X-ray AGN in 135 high mass clusters ($10^{14}M_{\odot}<M_{500}<4\times10^{15}M_{\odot}$) at $0.2<z<0.9$. It also agrees qualitatively with the results of \cite{Koulouridis2018} who examined the X-ray AGN fraction in 167 poor/intermediate richness clusters at $0.1<z<0.5$. While they found no suppression relative to the field in their overall sample, they also saw a suppression of cluster AGN activity when they selected only the highest mass clusters in their sample. 

This mass dependence of the cluster AGN fraction could help explain some of the differences between results in the literature. It could be that for poor and even intermediate richness clusters, neither environmental effects nor differences in galaxy-galaxy interactions are pronounced enough to manifest a substantial change in the cluster X-ray AGN fraction relative to the field (which will contain many unidentified groups). Indeed, studies that have investigated in detail the AGN population in modest overdensities, while controlling for stellar mass and AGN luminosity, have found no significant difference between the cluster and field AGN fraction \cite[e.g.][]{Yang2018,Powell2018}. Only studies that have probed the richest, most massive clusters ($M_{500}\gtrsim5\times10^{14}M_{\odot}$) have found a suppression of cluster AGN activity relative to the field \citep[e.g.][]{Martini2009,Ehlert2014}, although not all such studies have seen this \citep{Haggard2010}. 

\subsection{How does environment impact AGN activity?}

While we have observed a significant dependence of AGN activity on environment, wherein luminous (non central) galaxies in the most clusters are less likely to host an X-ray AGN, there could be many factors driving this dependence. Below we discuss a few possible explanations for this observed behaviour.

\subsubsection{Galaxy-cluster interactions}

The cluster environment is diverse, with ram-pressure stripping, harassment, and mergers being commonly observed in cluster outskirts, whereas tidal interactions with the cluster potential, evaporation, and starvation become more prominent near the cluster core \citep[see][]{Treu2003}. 
Looking at the distribution of AGN in clusters {\it relative to the inactive member population} can help disentangle which environmental effects play an important role in AGN activity.

For instance, differences in the phase-space distribution (see \cref{fig:local_density}) of active and inactive cluster members could hint toward environmental triggers of AGN activity. \cite{Haines2012} looked at this phase-space distribution of the X-ray AGN population in a sample of 26 clusters at $0.15<z<0.30$ and found that, relative to the inactive cluster member population, AGN tend to avoid regions with the lowest cluster-centric radii and relative velocities ($r<0.4r_{500}$ \& $|cz-\overline{cz}|/\sigma_{cz}$<0.8; dashed black region in the right panel of \cref{fig:local_density}). They found that cluster AGN preferentially lie along caustics with $|cz-\overline{cz}|/\sigma_{cz}\gtrsim1$, suggestive of AGN triggering on infall into the cluster. \cite{Ehlert2014} also found a  clear (factor $\sim 3$) suppression of the active galaxy fraction in the central regions ($r<r_{500}$) of massive clusters but little or no suppression in the cluster outskirts.
 
\begin{figure}
\begin{center}
\includegraphics[width=0.49\textwidth]{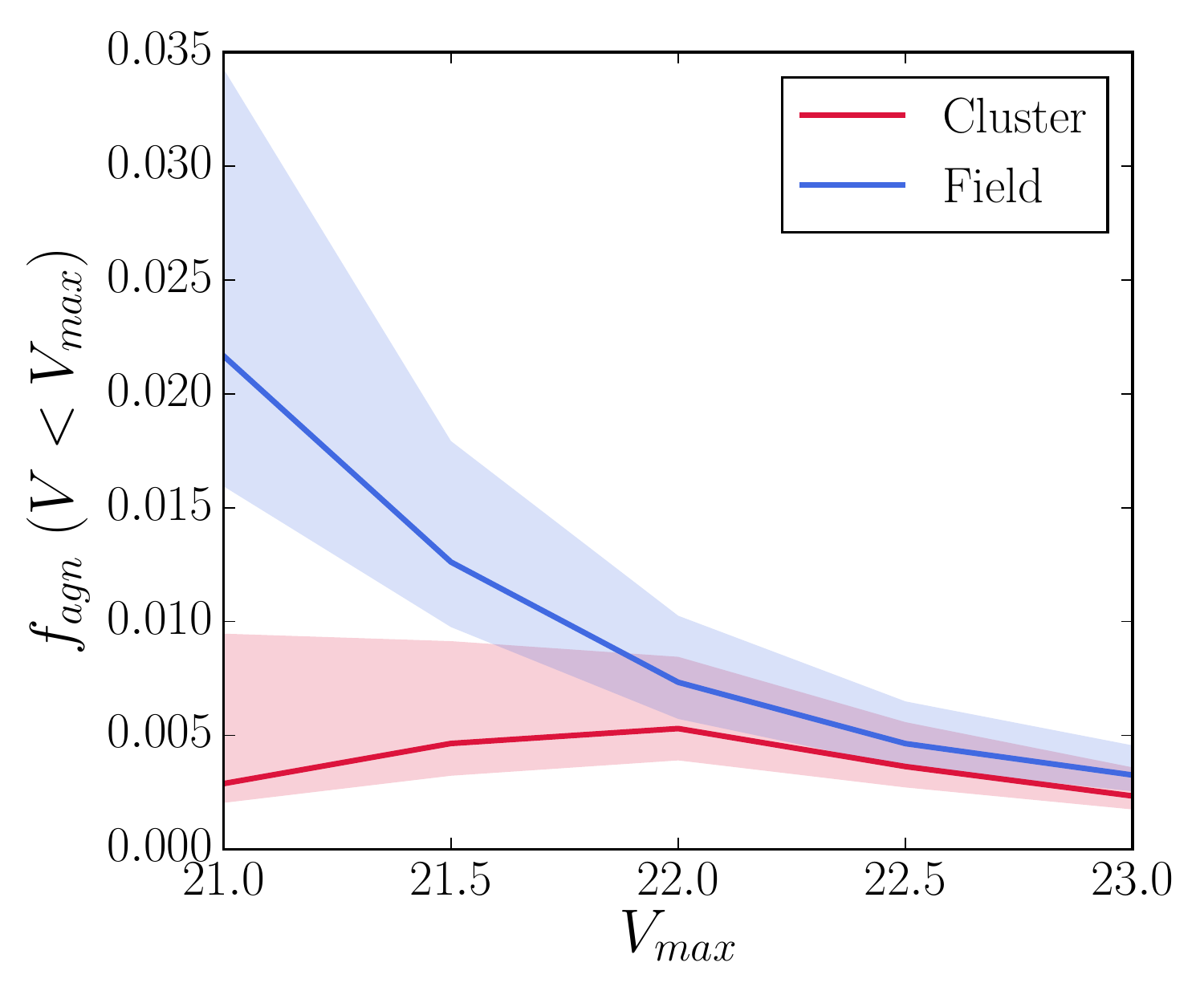}
\caption{The cluster AGN fraction is compared to that of the field as a function of $V$-band magnitude cut. 1$\sigma$ binomial error contours are shown.
Both cluster and field AGN are restricted to $L_X>6.8\times10^{42}~\mathrm{erg~s^{-1}}$ and the field population is constrained to $0.35<z<0.45$ to match the cluster redshifts.
A suppression of the cluster AGN fraction is observed only in the brightest, most massive galaxies.
}  \label{fig:fagn_vmax}
\end{center}
\end{figure}

While we see no significant difference between the distributions of active and inactive galaxies for the integrated population of with magnitudes $V<23$,  for the brightest galaxies with $V<21.5$ a clear suppression is observed (\cref{fig:fagn_vmax}). This is in agreement with \cite{Silverman2009} and \cite{Lopes2017} who found a suppression of the cluster AGN fraction relative to the field only for the most massive galaxies in their sample. Since the most massive galaxies in the field are typically the central galaxies of galaxy groups, while galaxies of similar mass in the cluster environment (having excluded BCGs) are satellites, this observed suppression could be due to differences between central and satellite populations. Within clusters, environmental effects such as ram pressure can efficiently strip satellite galaxies of their cool gas content, which is then channeled toward the central galaxy by gravity.
Qualitatively, this agrees with observations of large reservoirs of cold gas found in BCGs \citep[e.g.][]{Salome2008,McNamara2014,Russell2014} and the enhanced fraction of X-ray AGN found in BCGs relative to satellites \citep[i.e.][]{LYang2018}.

\subsubsection{Galaxy-galaxy interactions}

The dense environment of galaxy clusters changes the nature of galaxy-galaxy interactions relative to the field. In particular, the frequency of mergers and tidal interactions between galaxies in massive clusters is expected to scale inversely with cluster mass as $\sim M^{-1}$ as discussed in \cref{sec:massdep}. Our observation of the space density of cluster AGN scaling as $\sim M^{-2.0^{+0.8}_{-0.9}}$ is marginally steeper but consistent with this value. We note that one of target clusters, MACSJ0451$+$0006, hosts four of the seven cluster member AGN identified in this study. Excluding this cluster from the analysis, we find $\Phi\sim M^{-1.6^{+1.0}_{-1.1}}$ which is again consistent with an inverse mass dependence.

We visually test for galaxy-galaxy interactions in our cluster AGN and find that only one source has an unusual overdensity of neighbour galaxies. Additionally, if AGN activity in clusters is driven by interactions with other members, we might expect them to reside in regions of higher density than their inactive counterparts. However, \cref{fig:local_density} shows that this is not the case for our sample. Our results are in agreement with \cite{Pimbblet2013} who suggested that if AGN are being triggered by close encounters, any enhancement in local density must be washed out on shorter timescales than required for the AGN to turn on. \cite{Schawinski2007} found that it may take $\sim100$ Myrs for an AGN to build up its accretion disk and ``activate'' after any large-scale triggering event. Given the $\sigma_{cz}\sim1000$ km/s velocity dispersions of our clusters, galaxies could move $\sim100$ kpc in this time. Since the radii of circles enclosing the 10 nearest neighbors of cluster member galaxies range from 25 kpc to 100 kpc (see \cref{sec:density}), this displacement is sufficient to significantly dilute any signature of merger induced triggering in measures of local density. Future work utilizing a larger parent catalog and HST imaging will explore this question in more detail (Noordeh et al. in prep).

\subsubsection{Evidence for excess obscuration?}
Our observation that the cluster AGN number density falls with cluster mass could in principle be affected by obscuration effects. Recent studies have found that obscured AGN are more likely to reside in denser environments than unobscured AGN, even when controlling for luminosity, redshift, stellar mass, and Eddington ratio \citep{Powell2018,Mo2018}. We test this by comparing the X-ray hardness ratio (HR) for our cluster member AGN and COSMOS field comparison sample as
\begin{align}
HR = (H-S)/(H+S)
\end{align}
where $H$ is the net counts in the 2-8 keV band and $S$ is the net counts in the $0.5-2$ keV band. The hardness ratios for our cluster member AGN are tabulated in \cref{tab:agn}. If our sample of cluster member AGN were indeed subject to enhanced obscuration relative to the field we would expect this be reflected in larger HRs relative to the field. We compare the cluster and the field HR distributions with a two-sample Kolmogorov-Smirnov test and find $P=0.95$, indicating there is no evidence for enhanced obscuration of cluster member AGN in our sample. 

\subsection{MACS J0451.9$+$0006: Dynamically driven AGN triggering?} \label{sec:macs0451}

We note that one of our clusters, MACS J0451$+$0006, is host to four X-ray AGN while the rest of the clusters host either zero or one. This could be due to the relatively large galaxy density at large radii in MACS J0451$+$0006 (see \cref{fig:galdensity}) or statistical scatter across the sample. 
However, it is also possible that unique cluster properties are driving enhanced AGN triggering. 

\begin{figure}
\begin{center}
\includegraphics[width=0.5\textwidth]{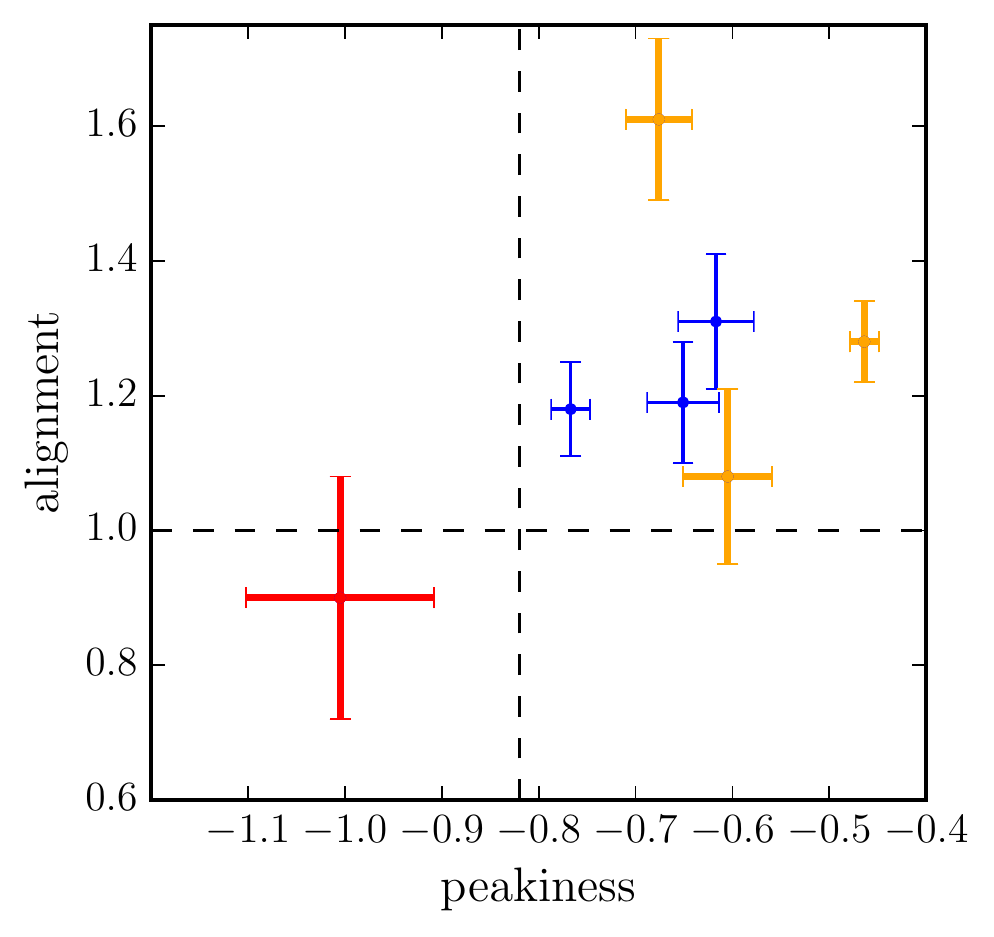}
\caption{The peakiness and alignment ``SPA'' metrics are shown for our cluster sample. MACS J0451.9+0006, host to four AGN, is shown in red and is the most dynamically disturbed cluster in our sample. Clusters that host a single AGN are marked in orange while those that host zero AGN are shown in blue. Dashed lines show the peakiness and alignment thresholds above which clusters are classified as relaxed (although not all of our clusters in this regime are classified as relaxed since ``symmetry'' must also be taken into account; see \protect\cite{Mantz2015}).
} \label{fig:macs0451_spa}
\end{center}
\end{figure}

To investigate this, we use the Symmetry, Peakiness, and Alignment (SPA) metrics of \cite{Mantz2015}. These are strong diagnostics of cluster dynamical activity and can be used to distinguish between relaxed and unrelaxed clusters. In particular, we plot the peakiness and alignment of our cluster sample in \cref{fig:macs0451_spa}. MACS J0451$+$0006 is marked in red. Based on all three SPA metrics it is the most dynamically disturbed cluster in our sample. Moreover, it has some of the most elliptical isophotes in a sample of 360 cluster observations from the ROSAT and {\it Chandra} archives \citep{Mantz2015}. In combination, this information is suggestive of a dramatic, head-on cluster merger, in agreement with a combined optical + X-ray analysis of the cluster morphology \citep{Mann2012}. This is particularly interesting since it has been suggested that cluster dynamical activity can trigger both SF and AGN activity in member galaxies \citep[e.g.][]{Sobral2015, Stroe2015}.
This can be due to merger driven shocks in the ICM inducing gas instabilities in cluster members \citep{Sobral2015}, by ram-pressure driven galaxy gas perturbations, or by enhanced galaxy-galaxy interactions \citep[e.g.][]{Canning2012}.

\begin{figure}
\begin{center}
\includegraphics[width=0.5\textwidth]{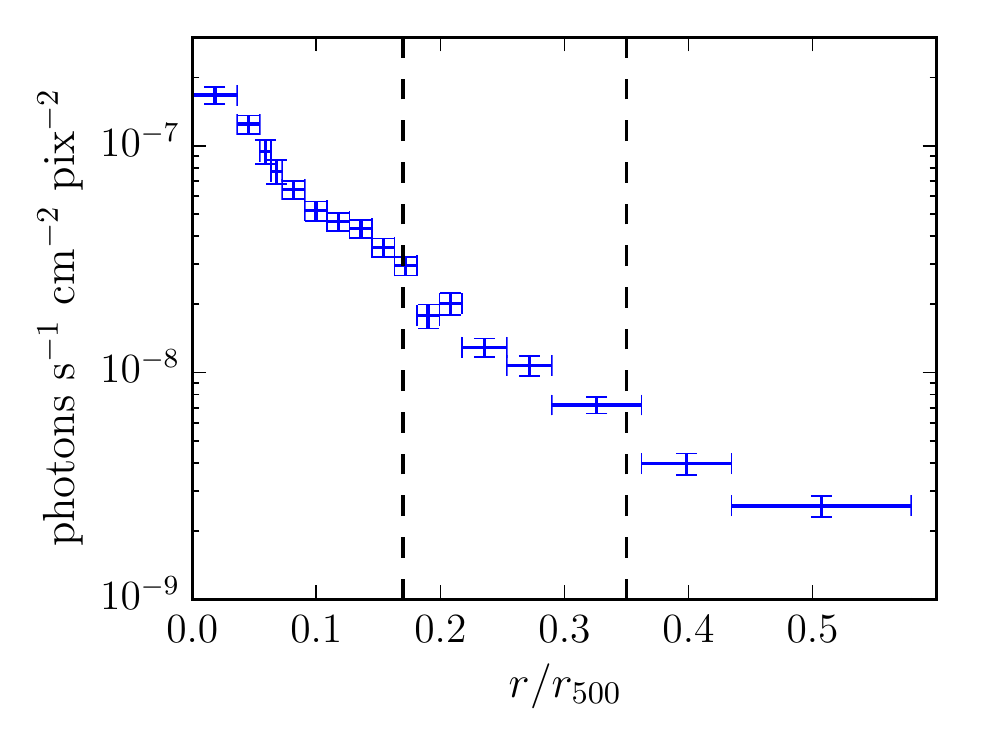}
\caption{The net X-ray surface brightness profile of MACS$~$J0451+0006 is shown as a function of cluster-centric radius. This utilizes the existing 27 ks of clean {\it Chandra} data available for the cluster. Each radial bin in the profile has a signal-to-noise ratio of at least eight. The dashed black lines show the radii at which the two central cluster member AGN reside. There is a hint of a discontinuity in the surface brightness profile slightly outside the location of the central-most AGN at $r=0.17r_{500}$, suggestive of a density discontinuity.}
\label{fig:macs0451_sb}
\end{center}
\end{figure}

Additionally, we look at the net X-ray surface brightness profile from the existing 27 ks of {\it Chandra} data shown in \cref{fig:macs0451_sb}. There are signs of a discontinuity in the surface brightness profile just past the location of the central-most AGN at $r = 0.17r_{500}$, perhaps a tantalizing hint of a passing shock in the ICM. Such shocks are expected to pass through galaxies on timescales of 10 - 50 Myr \citep{Sobral2015} and can induce perturbations in the IGM leading to the transport of gas to fuel AGN activity. The statistical significance of the discontinuity is modest and additional data is needed to confirm any shock in the ICM.

While larger samples are needed to confirm any dependence of AGN activity on cluster dynamical state, our study provides a clue that disturbed cluster environments may contribute to enhanced AGN triggering.

\section{Conclusions}

This study has analyzed the X-ray AGN population in seven massive galaxy clusters with $M>5\times10^{14}~M_\odot$ and  $0.35<z<0.45$. We probe all AGN with $L_X(0.5-8~\mathrm{keV})>6.8\times10^{42}~\mathrm{erg~s^{-1}}$ out to $2r_{500}$ in each cluster using a combination of {\it Chandra} imaging and VIMOS spectroscopy. We spectroscopically confirm the cluster membership of seven AGN and compare them both to the inactive cluster galaxy population and the field population from COSMOS. Our findings are as follows:

\begin{enumerate}
\item The cluster AGN fraction has a strong dependence on cluster mass with the space density of cluster AGN scaling as $\sim M^{-2.0^{+0.8}_{-0.9}}$. This result rules out zero mass dependence at the $2.5\sigma$ level.
\item We find that the cluster AGN fraction is suppressed relative to the field only for the optically brightest galaxies with $V<21.5$. This may be a consequence of a larger fraction of massive galaxies in the field being the centrals of their groups (sinks for gaseous fuel from satellites) relative to those in clusters.
\item Comparing the X-ray hardness ratio distributions of our cluster member AGN to an identically selected field sample, we find no evidence of enhanced X-ray obscuration of cluster members. 
%\item 
\item The most dynamically active cluster in our sample is host to four member AGN, as opposed to zero or one AGN in the other clusters. This hints towards cluster dynamical activity possibly playing a role in AGN triggering either by perturbing galactic gas supplies or enhancing galaxy-galaxy interactions.
\end{enumerate}

This study is limited by small number statistics due to the sparsity of cluster member AGN and the need for spectroscopic member confirmation. Furthermore, while we have discussed our results in the context of previous studies reported in the literature, we note that such comparisons are often complicated by differences in both AGN and cluster selection and the treatment of incompleteness. Larger studies with consistent, unbiased selection of AGN in massive clusters and in the field are needed to extend this work to higher precision. This could be accomplished by leveraging photometric redshifts to greatly expand sample sizes while minimizing contamination from interlopers. Additionally, the eROSITA all sky X-ray survey \citep{eRosita} combined with the Spectroscopic Identification of eROSITA Sources (SPIDERS) survey and 4MOST follow-up of galaxy groups and clusters, should yield spectra of $\sim40000$ X-ray AGN at $z<1$, unveiling the environmental dependence of AGN activity with exceptional statistical power.
To study the evolution of this environmental dependence out to the highest redshifts, however, we will require future X-ray observatories such as Athena and Lynx \citep{Athena2013,Lynx2019} and large optical telescopes \citep{TMT2015,ELTMOS2013} to robustly characterize high-redshift galaxy clusters and identify their AGN populations.

\section*{Acknowledgements}
The authors thank the anonymous referee for helpful feedback that improved the quality of this article. Support for this work was provided by NASA grant NNX16AL70G through the Astrophysics Data Analysis Program. EN acknowledges support from the Natural Sciences and Engineering Research Council of Canada PGS-D fellowship (516693). This paper is based on observations collected at the European Southern Observatory under ESO programmes 090.A-0958(B), 092.A-0405(A), and 094.A-0557(A). This work was supported in part by the U.S. Department of Energy under contract number DE-AC02-76SF00515. YQX acknowledges support from NSFC-11890693 \& 11421303, the CAS Frontier Science Key Research Program (QYZDJ-SSW-SLH006), and K.C. Wong Education Foundation.
We thank the DEEP2 team for providing their spec1D code as open access.
This research made use of Astropy,\footnote{http://www.astropy.org} a community-developed core Python package for Astronomy \citep{astropy:2013, astropy:2018}.

{\it
\section*{Data Availability}
The data underlying this article are available in the article and in its online supplementary material.
}

%%%%%%%%%%%%%%%%%%%%%%%%%%%%%%%%%%%%%%%%%%%%%%%%%%

%%%%%%%%%%%%%%%%%%%% REFERENCES %%%%%%%%%%%%%%%%%%

% The best way to enter references is to use BibTeX:
\bibliographystyle{mnras}
\bibliography{bibtex} % if your bibtex file is called example.bib

% Alternatively you could enter them by hand, like this:
% This method is tedious and prone to error if you have lots of references
% \begin{thebibliography}{99}
% \bibitem[\protect\citeauthoryear{Author}{2012}]{Author2012}
% Author A.~N., 2013, Journal of Improbable Astronomy, 1, 1
% \bibitem[\protect\citeauthoryear{Others}{2013}]{Others2013}
% Others S., 2012, Journal of Interesting Stuff, 17, 198
% \end{thebibliography}

%%%%%%%%%%%%%%%%%%%%%%%%%%%%%%%%%%%%%%%%%%%%%%%%%%

%%%%%%%%%%%%%%%%% APPENDICES %%%%%%%%%%%%%%%%%%%%%
%
 \appendix

 \section{Spectra of cluster member AGN} \label{app:spectra}

\begin{figure*} 
\begin{center}
\includegraphics[width=0.98\textwidth]{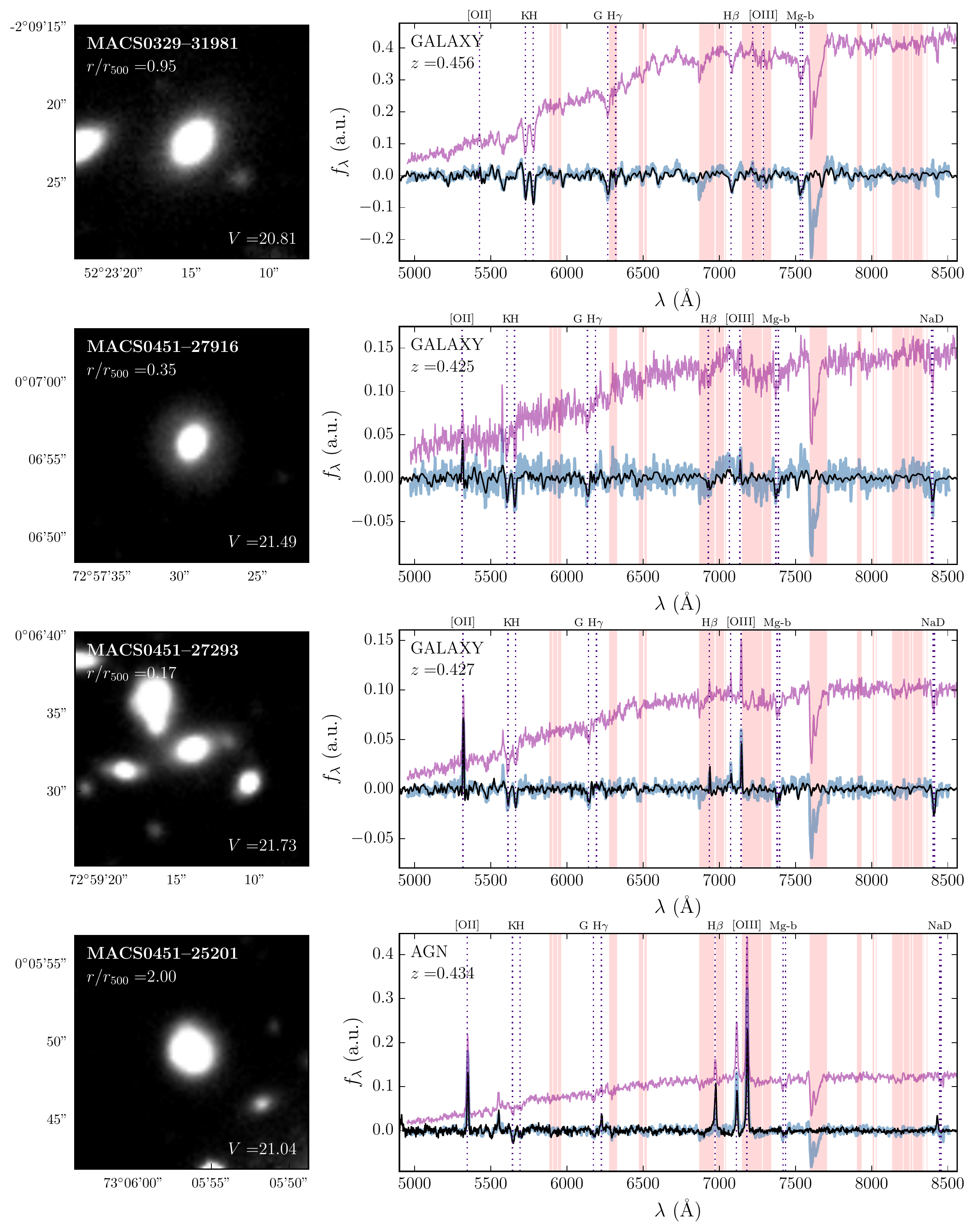}
\vspace{-1em}
\caption{{\bf Left:} $V$-band Subaru imaging $15''\times15''$ cutouts of X-ray point sources that are confirmed cluster member AGN. {\bf Right:} VIMOS Spectra are shown for the respective sources on the left. The raw 1D extracted spectrum is shown in purple, the continuum subtracted spectrum in blue, and the best fitting PCA reconstruction shown in black. Telluric absorption bands that were masked in the fitting procedure are shaded red. 
} \label{fig:cluster_agn_1}
\end{center}
\end{figure*}

\begin{figure*} 
\ContinuedFloat
\begin{center}
\includegraphics[width=0.98\textwidth]{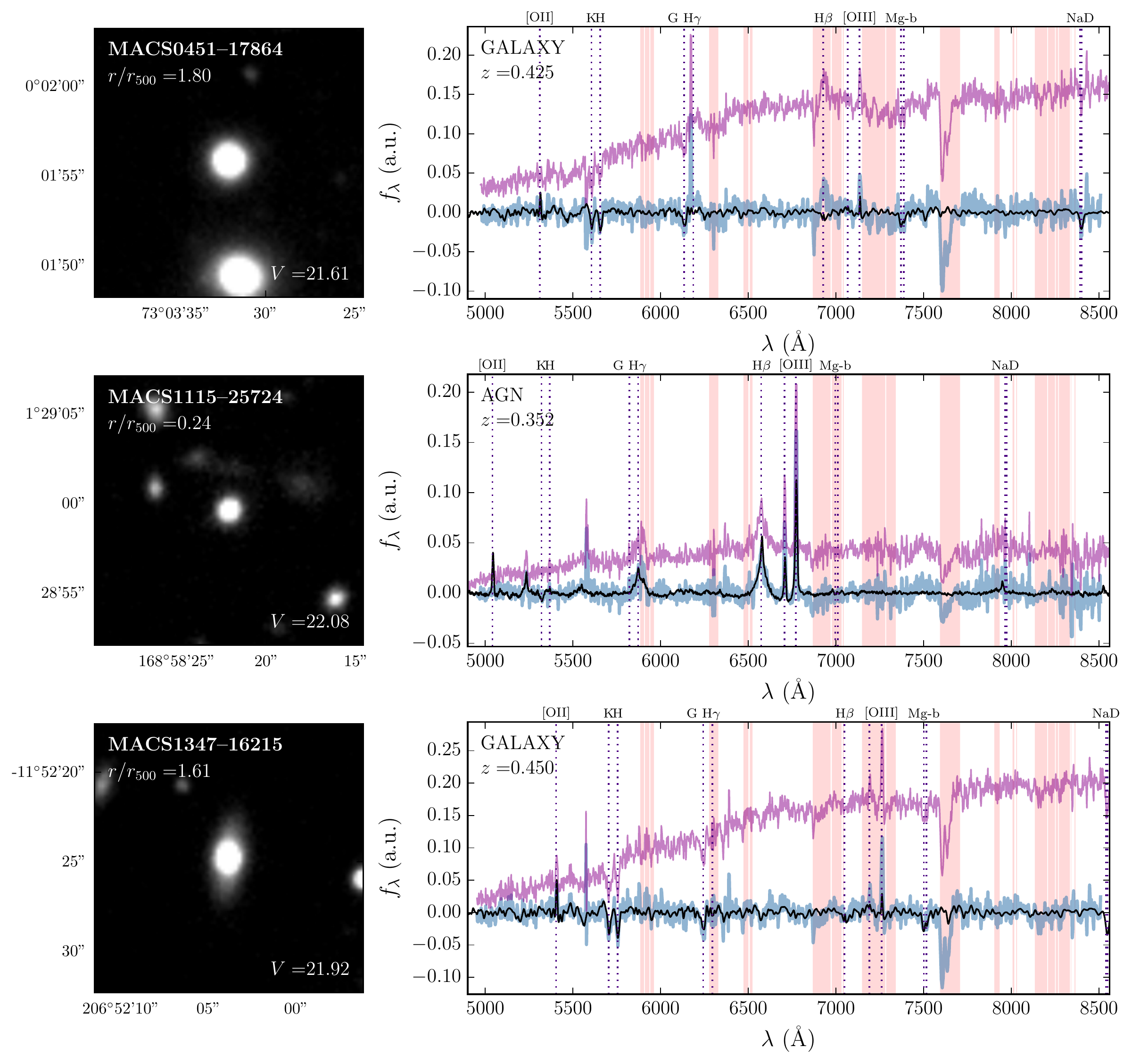}
\caption{continued.}
 \label{fig:cluster_agn_2}
\end{center}
\end{figure*}

% If you want to present additional material which would interrupt the flow of the main paper,
% it can be placed in an Appendix which appears after the list of references.

%%%%%%%%%%%%%%%%%%%%%%%%%%%%%%%%%%%%%%%%%%%%%%%%%%

% Don't change these lines
\bsp	% typesetting comment
\label{lastpage}
\end{document}